\RequirePackage{lineno}

\documentclass[aps,prl,twocolumn,superscriptaddress,showpacs,floatfix]{revtex4}

\usepackage{epsfig}
\usepackage{psfrag}
\usepackage{graphicx}
\usepackage{dcolumn}
\usepackage{bm}
\usepackage{epsfig}
\usepackage{lineno}
\usepackage{multirow}
\usepackage{slashed}
\usepackage{rotating}
\usepackage{subfig}
\usepackage{caption}
\usepackage{verbatim} 
\usepackage{mathrsfs}

\newcommand{\pt}{\mbox{$p_{T}$}~}
\newcommand{\pte}{$p_{T}$} 
\newcommand{\mjl}{\mbox{$D_{\text{MJL}}$}~}
\newcommand{\mjle}{\mbox{$D_{\text{MJL}}$}}

\newcommand{\dzero} {\mbox{D0}~}

\newcommand{\GeV} {\ensuremath{\mathrm{Ge\kern -0.1em V}}~}
\newcommand{\GeVe} {\ensuremath{\mathrm{Ge\kern -0.1em V}}}
\newcommand{\TeV} {\ensuremath{\mathrm{Te\kern -0.1em V}}}

\newcommand{\ppbar}{\mbox{$p\overline{p}$}~}

\newcommand{\ttbar}{\mbox{$t\overline{t}$}}

\newcommand{\alpgen}{{\sc alpgen}~}
\newcommand{\alpgene}{{\sc alpgen}}
\newcommand{\pythia}{{\sc pythia}~}
\newcommand{\pythiae}{{\sc pythia}}
\newcommand{\mcfm}{{\sc mcfm}~}

\newcommand{\sherpa}{{\sc sherpa}~}
\newcommand{\sherpae}{{\sc sherpa}}

\def\gsim{\mathrel{\rlap{\raise.4ex\hbox{$>$}} {\lower.6ex\hbox{$\sim$}}}}
\def\lsim{\mathrel{\rlap{\raise.4ex\hbox{$<$}} {\lower.6ex\hbox{$\sim$}}}}

\newcommand{\ptj}{\mbox{$p_{T}^{\text{jet}}$~}}
\newcommand{\ptz}{\mbox{$p_{T}^{Z}$~}}
\newcommand{\ptje}{\mbox{$p_{T}^{\text{jet}}$}}
\newcommand{\ptze}{\mbox{$p_{T}^{Z}$}}
\newcommand{\etaj}{\mbox{$\eta^{\text{jet}}$~}}
\newcommand{\etaje}{\mbox{$\eta^{\text{jet}}$}}

\newcommand{\ZcZj}{\mbox{$R_{c/\text{jet}}$}}
\newcommand{\ZcZb}{\mbox{$R_{c/b}$}}

\newcommand{\bl} {MVA$_{bl}$~}

\begin{document}

\hspace{5.2in} \mbox{FERMILAB-PUB-13-329-E}

\title{\boldmath Measurement of associated production of $Z$ bosons with charm quark jets\\
in $p\bar{p}$ collisions at $\sqrt s=1.96~\TeV$}

\affiliation{LAFEX, Centro Brasileiro de Pesquisas F\'{i}sicas, Rio de Janeiro, Brazil}
\affiliation{Universidade do Estado do Rio de Janeiro, Rio de Janeiro, Brazil}
\affiliation{Universidade Federal do ABC, Santo Andr\'e, Brazil}
\affiliation{University of Science and Technology of China, Hefei, People's Republic of China}
\affiliation{Universidad de los Andes, Bogot\'a, Colombia}
\affiliation{Charles University, Faculty of Mathematics and Physics, Center for Particle Physics, Prague, Czech Republic}
\affiliation{Czech Technical University in Prague, Prague, Czech Republic}
\affiliation{Institute of Physics, Academy of Sciences of the Czech Republic, Prague, Czech Republic}
\affiliation{Universidad San Francisco de Quito, Quito, Ecuador}
\affiliation{LPC, Universit\'e Blaise Pascal, CNRS/IN2P3, Clermont, France}
\affiliation{LPSC, Universit\'e Joseph Fourier Grenoble 1, CNRS/IN2P3, Institut National Polytechnique de Grenoble, Grenoble, France}
\affiliation{CPPM, Aix-Marseille Universit\'e, CNRS/IN2P3, Marseille, France}
\affiliation{LAL, Universit\'e Paris-Sud, CNRS/IN2P3, Orsay, France}
\affiliation{LPNHE, Universit\'es Paris VI and VII, CNRS/IN2P3, Paris, France}
\affiliation{CEA, Irfu, SPP, Saclay, France}
\affiliation{IPHC, Universit\'e de Strasbourg, CNRS/IN2P3, Strasbourg, France}
\affiliation{IPNL, Universit\'e Lyon 1, CNRS/IN2P3, Villeurbanne, France and Universit\'e de Lyon, Lyon, France}
\affiliation{III. Physikalisches Institut A, RWTH Aachen University, Aachen, Germany}
\affiliation{Physikalisches Institut, Universit\"at Freiburg, Freiburg, Germany}
\affiliation{II. Physikalisches Institut, Georg-August-Universit\"at G\"ottingen, G\"ottingen, Germany}
\affiliation{Institut f\"ur Physik, Universit\"at Mainz, Mainz, Germany}
\affiliation{Ludwig-Maximilians-Universit\"at M\"unchen, M\"unchen, Germany}
\affiliation{Panjab University, Chandigarh, India}
\affiliation{Delhi University, Delhi, India}
\affiliation{Tata Institute of Fundamental Research, Mumbai, India}
\affiliation{University College Dublin, Dublin, Ireland}
\affiliation{Korea Detector Laboratory, Korea University, Seoul, Korea}
\affiliation{CINVESTAV, Mexico City, Mexico}
\affiliation{Nikhef, Science Park, Amsterdam, the Netherlands}
\affiliation{Radboud University Nijmegen, Nijmegen, the Netherlands}
\affiliation{Joint Institute for Nuclear Research, Dubna, Russia}
\affiliation{Institute for Theoretical and Experimental Physics, Moscow, Russia}
\affiliation{Moscow State University, Moscow, Russia}
\affiliation{Institute for High Energy Physics, Protvino, Russia}
\affiliation{Petersburg Nuclear Physics Institute, St. Petersburg, Russia}
\affiliation{Instituci\'{o} Catalana de Recerca i Estudis Avan\c{c}ats (ICREA) and Institut de F\'{i}sica d'Altes Energies (IFAE), Barcelona, Spain}
\affiliation{Uppsala University, Uppsala, Sweden}
\affiliation{Lancaster University, Lancaster LA1 4YB, United Kingdom}
\affiliation{Imperial College London, London SW7 2AZ, United Kingdom}
\affiliation{The University of Manchester, Manchester M13 9PL, United Kingdom}
\affiliation{University of Arizona, Tucson, Arizona 85721, USA}
\affiliation{University of California Riverside, Riverside, California 92521, USA}
\affiliation{Florida State University, Tallahassee, Florida 32306, USA}
\affiliation{Fermi National Accelerator Laboratory, Batavia, Illinois 60510, USA}
\affiliation{University of Illinois at Chicago, Chicago, Illinois 60607, USA}
\affiliation{Northern Illinois University, DeKalb, Illinois 60115, USA}
\affiliation{Northwestern University, Evanston, Illinois 60208, USA}
\affiliation{Indiana University, Bloomington, Indiana 47405, USA}
\affiliation{Purdue University Calumet, Hammond, Indiana 46323, USA}
\affiliation{University of Notre Dame, Notre Dame, Indiana 46556, USA}
\affiliation{Iowa State University, Ames, Iowa 50011, USA}
\affiliation{University of Kansas, Lawrence, Kansas 66045, USA}
\affiliation{Louisiana Tech University, Ruston, Louisiana 71272, USA}
\affiliation{Northeastern University, Boston, Massachusetts 02115, USA}
\affiliation{University of Michigan, Ann Arbor, Michigan 48109, USA}
\affiliation{Michigan State University, East Lansing, Michigan 48824, USA}
\affiliation{University of Mississippi, University, Mississippi 38677, USA}
\affiliation{University of Nebraska, Lincoln, Nebraska 68588, USA}
\affiliation{Rutgers University, Piscataway, New Jersey 08855, USA}
\affiliation{Princeton University, Princeton, New Jersey 08544, USA}
\affiliation{State University of New York, Buffalo, New York 14260, USA}
\affiliation{University of Rochester, Rochester, New York 14627, USA}
\affiliation{State University of New York, Stony Brook, New York 11794, USA}
\affiliation{Brookhaven National Laboratory, Upton, New York 11973, USA}
\affiliation{Langston University, Langston, Oklahoma 73050, USA}
\affiliation{University of Oklahoma, Norman, Oklahoma 73019, USA}
\affiliation{Oklahoma State University, Stillwater, Oklahoma 74078, USA}
\affiliation{Brown University, Providence, Rhode Island 02912, USA}
\affiliation{University of Texas, Arlington, Texas 76019, USA}
\affiliation{Southern Methodist University, Dallas, Texas 75275, USA}
\affiliation{Rice University, Houston, Texas 77005, USA}
\affiliation{University of Virginia, Charlottesville, Virginia 22904, USA}
\affiliation{University of Washington, Seattle, Washington 98195, USA}
\author{V.M.~Abazov} \affiliation{Joint Institute for Nuclear Research, Dubna, Russia}
\author{B.~Abbott} \affiliation{University of Oklahoma, Norman, Oklahoma 73019, USA}
\author{B.S.~Acharya} \affiliation{Tata Institute of Fundamental Research, Mumbai, India}
\author{M.~Adams} \affiliation{University of Illinois at Chicago, Chicago, Illinois 60607, USA}
\author{T.~Adams} \affiliation{Florida State University, Tallahassee, Florida 32306, USA}
\author{J.P.~Agnew} \affiliation{The University of Manchester, Manchester M13 9PL, United Kingdom}
\author{G.D.~Alexeev} \affiliation{Joint Institute for Nuclear Research, Dubna, Russia}
\author{G.~Alkhazov} \affiliation{Petersburg Nuclear Physics Institute, St. Petersburg, Russia}
\author{A.~Alton$^{a}$} \affiliation{University of Michigan, Ann Arbor, Michigan 48109, USA}
\author{A.~Askew} \affiliation{Florida State University, Tallahassee, Florida 32306, USA}
% ...
%
%
\collaboration{The D0 Collaboration\footnote{with visitors from
$^{a}$Augustana College, Sioux Falls, SD, USA,
$^{b}$The University of Liverpool, Liverpool, UK,
$^{c}$DESY, Hamburg, Germany,
$^{d}$Universidad Michoacana de San Nicolas de Hidalgo, Morelia, Mexico
$^{e}$SLAC, Menlo Park, CA, USA,
$^{f}$University College London, London, UK,
$^{g}$Centro de Investigacion en Computacion - IPN, Mexico City, Mexico,
$^{h}$Universidade Estadual Paulista, S\~ao Paulo, Brazil,
$^{i}$Karlsruher Institut f\"ur Technologie (KIT) - Steinbuch Centre for Computing (SCC)
and
$^{j}$Office of Science, U.S. Department of Energy, Washington, D.C. 20585, USA.
}} \noaffiliation
\vskip 0.25cm

\date{August 20, 2013}

\begin{abstract}  
We present the first measurements of the ratios of cross sections,
$\sigma(p\bar{p}\rightarrow Z+c~\text{jet})$/$\sigma(p\bar{p}\rightarrow Z+\text{jet})$
and $\sigma(p\bar{p}\rightarrow Z+c~\text{jet})$/$\sigma(p\bar{p}\rightarrow Z+b~\text{jet})$
for the associated production of a $Z$ boson with at least one charm or bottom quark jet. 
Jets have transverse momentum $\ptje>20$ GeV and pseudorapidity $|\etaje|<2.5$. 
These cross section ratios are measured differentially as a function of jet and $Z$ boson transverse momenta, 
based on 9.7 fb$^{-1}$ of \ppbar collisions collected with the D0 detector at the Fermilab Tevatron 
Collider at $\sqrt{s} = 1.96$ TeV. 
The measurements show significant deviations from perturbative 
QCD calculations and predictions from various event generators.

\end{abstract}

\maketitle

\newpage

Studies of $Z$ boson production in association 
with heavy flavor (HF) jets originating from $b$ or $c$ quarks 
provide important tests of perturbative quantum chromodynamics 
(pQCD) calculations~\cite{Campbell}. 
A good theoretical description of 
these processes 
is essential since they form a major background for a variety of physics processes, including
standard model Higgs boson production in association with
a $Z$ boson, $ZH(H\rightarrow b\bar{b})$~\cite{zhllbb}.
Furthermore, the relative contributions of the different flavors to the background
is important since $Z+c~\text{jet}$ events can be misidentified as $Z+b~\text{jet}$ events, 
or vice versa, and therefore introduce additional uncertainties into measurements.

The ratio of $Z+b~\text{jet}$ to inclusive $Z+\text{jet}$ production cross sections
for events with one or more jets has previously been measured by the
CDF~\cite{CDFPaper,CDFPaperII}  and D0~\cite{Zb_PRL,Zb}  Collaborations. 
This Letter reports the first measurement of associated charm jet production 
with a $Z$ boson.
In particular, we present the measurement of the ratio of cross sections 
for $Z+c~\text{jet}$ to $Z+\text{jet}$ production
as well as $Z+c~\text{jet}$ to $Z+b~\text{jet}$ production in events with at least one jet.
The measurement of the ratio of cross sections benefits from the
cancellation of several systematic uncertainties and therefore allows for a more precise
comparison of data with the theoretical predictions.
These ratio measurements are also presented differentially as a function of
the transverse momenta of the jet (\ptje) and $Z$ boson (\ptze).

The current analysis is based on the complete Run~II
data sample collected using the D0 detector~\cite{d0det} at
Fermilab's Tevatron \ppbar Collider with a center-of-mass
energy of 1.96~TeV, and corresponds to an integrated luminosity of 9.7~fb$^{-1}$
following the application of relevant data quality requirements.
We use the same triggering, selections, object reconstruction, and event modeling as described in the recent D0 measurement 
of $Z + b~\text{jet}$ production~\cite{Zb}, but with a dedicated strategy for the extraction of the $c$-jet fraction.  
Events must contain a $Z \rightarrow \ell \ell$ candidate 
with a dilepton invariant mass in the range $70~<M_{\ell\ell}<110~\GeVe~(\ell = e,\mu)$. 

Dielectron ($ee$) events are required to have two electrons,  with no requirement on the sign 
of their electric charge, with transverse momentum \pte$>15~\GeVe$ 
identified through electromagnetic (EM) showers in the calorimeter. 
One electron must be identified in the central calorimeter (CC), within a 
pseudorapidity~\cite{def} region $|\eta|<1.1$, while the second electron can be reconstructed either 
in the CC or the endcap calorimeters, $1.5<|\eta|<2.5$. 
Dimuon ($\mu\mu$) events are required to have two oppositely charged 
muons, with \pte$>15~\GeVe$ and $|\eta|<2$, detected in the muon spectrometer 
and matched to central tracker tracks. 
In addition, at least one hadronic jet must be reconstructed in the event using an iterative midpoint cone algorithm~\cite{RunIIcone} 
with a cone size of $\Delta R =  \sqrt{(\Delta\varphi)^{2}+(\Delta y)^{2}}= 0.5$ where
$\varphi$ is the azimuthal angle and $y$ is the rapidity. This jet must satisfy
\ptj$>20~\GeVe$ and $|\etaj|<2.5$.

Several processes can mimic the signature of $Z+\text{jet}$ events.
These include top quark pair ($t\bar{t}$), diboson ($WW$, $WZ$, and $ZZ$), and multijet production.
To suppress the contributions from \ttbar~production, 
events with significant imbalance in the measured transverse 
energy, $\slashed{E}_{T}$, due to undetected neutrinos from the $W$ boson decay 
($t \rightarrow Wb \rightarrow \ell \nu_{\ell} b$), 
are rejected if $\slashed{E}_{T} > 60$~GeV.
These selection criteria retain an inclusive sample of 
176,498 $Z+\text{jet}$ event candidates in the $ee$ and $\mu\mu$ channels.

To estimate acceptances, efficiencies, and backgrounds, the
$Z+\text{jet}$ events (including HF jets) and $t\bar{t}$ events are modeled by
{\sc alpgen}~\cite{alpgen}, which generates sub-processes using higher-order 
QCD tree-level matrix elements (ME), interfaced with the
{\sc pythia} Monte Carlo (MC) event generator~\cite{pythia} for parton showering and hadronization
and {\sc evtgen}~\cite{evtgen} for modeling the decay of particles containing $b$ and $c$ quarks.
Inclusive diboson production is simulated with
{\sc pythia}. The {\sc cteq6l1}~\cite{cteq6} 
parton distribution functions (PDFs) are used in these simulations 
and the cross sections are scaled to the corresponding 
higher-order theoretical calculations. 
For the diboson and $Z+\text{jet}$ processes, 
including $Z + b\bar{b}$ and $Z + c\bar{c}$
production, next-to-leading order (NLO) cross section
predictions are taken from {\sc mcfm}~\cite{diboson}.
The $t\bar{t}$ cross section is determined from approximate next-to-NLO
calculations~\cite{ttbar}. 
To improve the modeling of the \pt distribution of the $Z$ boson, simulated 
$Z+\text{jet}$ events are also reweighted to be consistent with the measured \pt 
spectrum of $Z$ bosons observed in data~\cite{zpt}. 
The multijet background, where jets are  misidentified as leptons, 
is determined using a data-driven method, as described in 
the recent D0 publication~\cite{Zb}.
The fractions of non-$Z+\text{jet}$ events in the $ee$ and $\mu\mu$ samples 
are about 9.6\% and 1.3\%, respectively. 
These fractions are dominated by multijet production where a jet is either 
mis-reconstructed as a lepton in the electron channel, or a lepton from
decays of hadrons in a jet that passes the isolation requirement, in the muon channel. 

This analysis employs a two-step procedure to determine the HF content
of jets in the selected $Z+\text{jet}$ events. We employ a HF tagging algorithm~\cite{bid}
to enrich the sample in $b$ and $c$ jets.
The $b$, $c$, and light jet composition of the data is then extracted via a template-based fit. 

Jets considered for HF tagging are subject to a preselection requirement,
known as taggability~\cite{bid} to decouple the intrinsic performance 
of the HF jet tagging algorithm from effects related to track reconstruction efficiency. 
For this purpose, the jet is required to have at least two associated
tracks with \pte$>0.5~\GeVe$ and the highest-\pt track must have \pte$>1~\GeVe$. 
The efficiency of the taggability requirement is 90\% for both $c$ and $b$ jets.

The HF tagging algorithm is based on a multivariate analysis (MVA) technique~\cite{MVA} 
that provides an improved performance over the neural network HF tagging
discriminant, described in Ref.~\cite{bid}, used in earlier D0 analyses. 
This new algorithm, known as MVA$_{bl}$, 
also utilizes the relatively long lifetime of HF hadrons with respect to
their lighter counterparts. Events with at least one jet passing 
the HF tagging selection are considered in the analysis.

\begin{figure}
\centering
\includegraphics[width=0.49\columnwidth]{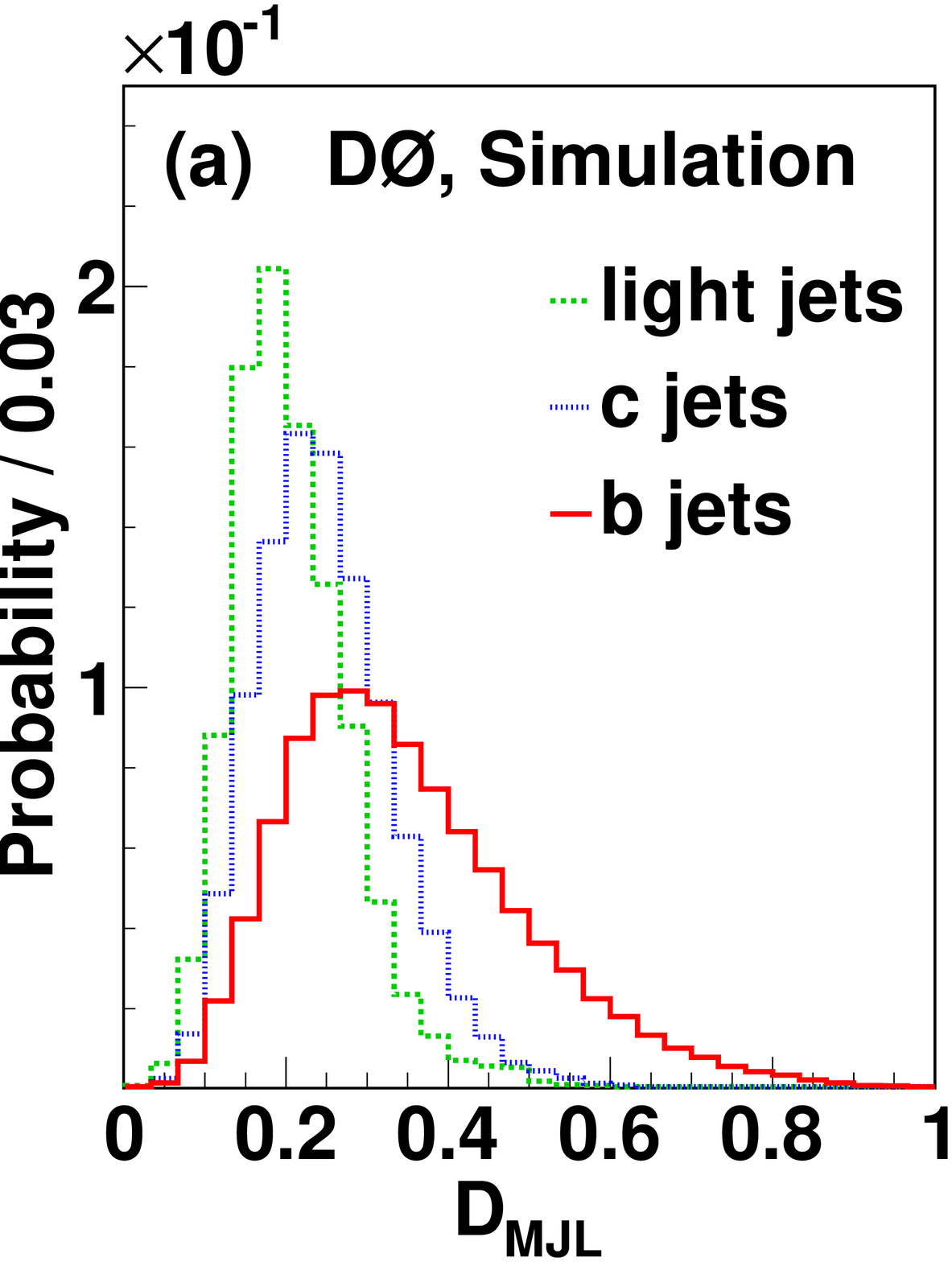} \includegraphics[width=0.49\columnwidth]{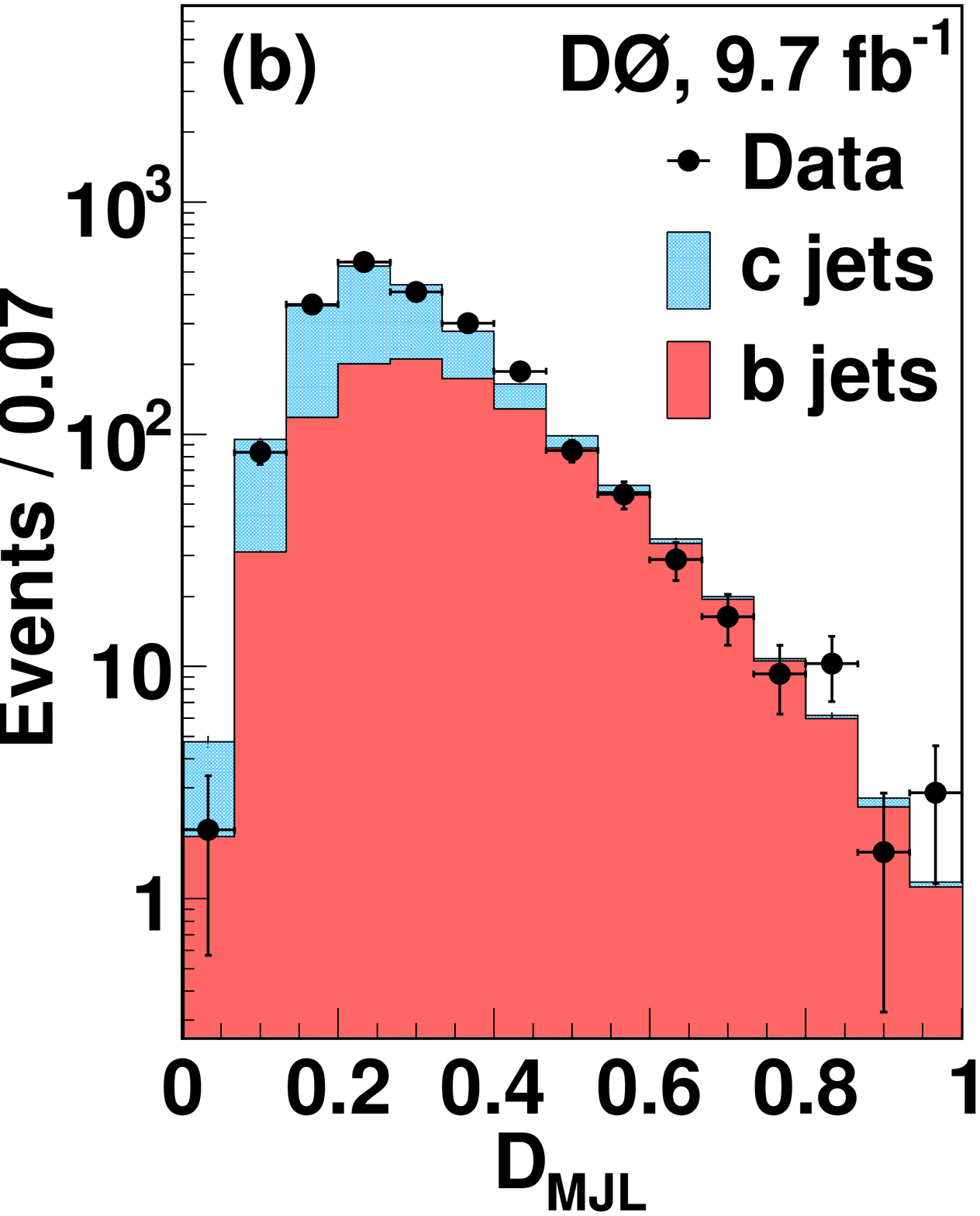}
\caption{\label{fig:channel_fit} (color online) 
(a) The probability densities of the \mjl discriminant for $b$, $c$,
and light jets passing the final selection requirements. These templates are
obtained from MC.
(b) The \mjl discriminant distribution of
events in the combined sample after background subtraction. 
The distributions of the $b$ and $c$
jets are weighted by the fractions found from the fit. Uncertainties
are statistical only.}
\end{figure}

To extract the fraction of different flavor jets in the data sample, 
a second discriminant, \mjle, is employed, which offers improved flavor separation for jets passing our \bl requirement~\cite{Zb}. 
It is a combination of two discriminating variables, the 
secondary vertex mass ($M_{\text{SV}}$) and the jet lifetime impact parameter (JLIP)~\cite{bid}:
$\mjl = 0.5\times(M_{\text{SV}}/5~\GeVe - \ln(\text{JLIP})/20)$.
The coefficients in this expression are chosen to optimize the separation 
of the HF and light quark components.
Fig.~\ref{fig:channel_fit}(a) shows the
\mjl distributions (templates) obtained from simulations of all three considered
jet flavors that pass an MVA$_{bl} > 0.5$ requirement.

To measure the relative fraction of $c$ jets in the HF enriched sample, 
the following two approaches were considered. 
The first is based on the methods used in Ref.~\cite{Zb} 
where the composition of $b$, $c$, and light jets is extracted by fitting MC templates to the data.  
This approach yields a large uncertainty on the $c$-jet 
fraction since the \mjl distributions of  $c$ and light jets are similar. 
The second approach is to suppress events with light jets by employing 
a more stringent \bl requirement. 
The remaining small $Z+\text{light jet}$ contribution, 
as estimated with data-corrected simulations, is then subtracted from the data.  
This allows for the data to be fit with only $b$ and $c$ jet templates.
Both methods yield consistent results, but the second method
benefits from a reduced overall uncertainty since
only the normalization of $b$ and $c$ jet templates are allowed to vary
when fitting the data.
Events are retained for further 
analysis if they contain at least one jet with an MVA$_{bl}$
output greater than 0.5.
After these requirements, 2,665 $Z+\text{jet}$ events are
selected where only the highest-\pt HF tagged jet is examined.
The efficiencies of the \bl selection for $b$ and $c$ jets, and the light jet misidentification rate
 are 40\%, 9.0\%, and 0.24\%, respectively. 
 The background is dominated by $Z+\text{light jet}$ events that comprise 12\% of the total sample.
Before the two parameter fit, all background components are
subtracted from the data, yielding a sample of 2,125 events.

We measure the fraction of events that contain at least one $b$ or $c$ jet 
in the $ee$ and $\mu\mu$ samples separately, yielding $c$ jet flavor fractions of 
$0.509\pm0.041\thinspace (\mbox{stat.})$ and $0.470\pm0.039 \thinspace(\mbox{stat.})$, respectively. 
Since these are consistent and the kinematics of the corresponding
events are similar, we combine the two samples to increase the statistical power
of the fit.
The combined \mjl distribution of the HF-enriched background subtracted data 
and the fitted templates for the $b$ and $c$ jets are shown in Fig.~\ref{fig:channel_fit}(b).
The corresponding fractions of $c$ and $b$ jets in the data are found to be 
$0.486\pm0.027\thinspace (\text{stat.})$ and $0.514\pm0.027\thinspace (\text{stat.})$, respectively.
These fractions are combined with the relevant 
detector acceptances and efficiencies  to determine the ratios
of cross sections using

\begin{eqnarray}
  \begin{array}{ccccccc}
R_{c/\text{jet}} & \equiv & \frac{\sigma(Z+c~\text{jet})}{\sigma(Z+\text{jet})} & = & \frac{N_{\text{HF}} f_{c}}{N_{\text{incl}}\,\epsilon_{tag}^{c}} & \times & \frac{\mathcal{A}_{\text{incl}}}{\mathcal{A}_{c}} \\
R_{c/b} & \equiv & \frac{\sigma(Z+c~\text{jet})}{\sigma(Z+b~\text{jet})} &= &\frac{f_{c}~\epsilon_{tag}^{b}}{f_{b}~\epsilon_{tag}^{c}} & \times & \frac{\mathcal{A}_{b}}{\mathcal{A}_{c}}
  \end{array}
\label{eq:zczb}
\end{eqnarray}

\noindent where  $N_{\text{incl}}$ is the total number of $Z+\text{jet}$ events
before the tagging requirements, $N_{\text{HF}}$ is the number of $Z+\text{jet}$
events used in the \mjl fit, $f_{b(c)}$ is the extracted $b(c)$ jet
fraction, and $\epsilon_{tag}^{b(c)}$ is the selection efficiency 
for $b(c)$ jets, which combines the efficiencies for taggability and MVA$_{bl}$
discriminant selection. $N_{\text{incl}}$ and $N_{\text{HF}}$
correspond to the number of events that remain after the contributions 
from various background processes have been subtracted. 
We subtract contributions from $t\bar{t}$, diboson, and multijet production to obtain 
$N_{\text{incl}}$, while we also subtract the $Z+\text{light jet}$ events when 
calculating $N_{\text{HF}}$.

The detector acceptances for the inclusive jet, $\mathcal{A}_{\text{incl}}$,
and $b (c)$ jets, $\mathcal{A}_{b (c)}$, are determined from MC simulation in the kinematic region that
satisfies the $p_T$ and $\eta$ requirements for leptons and jets. 
In these ratios, the effect of migration of events near the kinematic
thresholds, or between neighboring kinematic bins, due to detector
resolution is found to be negligible.

Using Eqs.~(\ref{eq:zczb}), the ratio of the cross sections $Z+c~\text{jet}$ to inclusive $Z+\text{jet}$ 
in the combined $\mu\mu$ and $ee$ channel,
\ZcZj, is $0.0829\pm0.0052 \thinspace (\text{stat.})$ and the ratio of cross sections
$Z+c~\text{jet}$ to $Z+b~\text{jet}$, \ZcZb, is found to be 
$4.00\pm0.21\thinspace (\text{stat.})$.
These ratios have also been measured differentially as  
a function of \ptj and \ptze.
For \ZcZj, the highest-\pt tagged jet from the 
HF enriched sample is used in the numerator,
while the denominator uses the highest-\pt jet from 
the $Z+\text{jet}$ sample.
The selected bin sizes along with the corresponding statistics of data events are listed
in Table \ref{tab:Final}. In each case, all the quantities that
enter into Eqs.~(\ref{eq:zczb}) are determined in each bin separately.

\begin{table}
\caption{\label{tab:Final} Summary of bins, data statistics and the measured ratios along with 
the statistical and systematic relative uncertainties in percent.
Bin centers, shown in parenthesis, are chosen using the prescription in Ref.~\cite{Terry}.}
\begin{tabular}{rccccccc}
\hline 
\multirow{2}{*}{\ptj[GeV]} & \multirow{2}{*}{$N$} & \multirow{2}{*}{\ZcZj} & Stat. & Syst.  & \multirow{2}{*}{\ZcZb} & Stat. & Syst. \tabularnewline
 &  &  & [\%] & [\%] &  & [\%] & [\%]  \tabularnewline
\hline 
$20-30\thinspace (24.6)$ & 741 & 0.068 & 12 & 16  & 3.64 & 8.5 & 21 \tabularnewline
$30-40\thinspace (34.3)$ & 525 & 0.084 & 11 & 12 & 3.97 & 8.3 & 14 \tabularnewline
$40-60\thinspace (47.3)$ & 474 & 0.099 & 11 & 9.1 & 3.98 & 10 & 13 \tabularnewline
$60-200\thinspace (78.0)$ & 380 & 0.085 & 13 & 11 & 4.30 & 13 & 14 \tabularnewline
\ptz[GeV] &  &  &  &  &  &  &   \tabularnewline
$0-20\thinspace (10.2)$ & 285 & 0.041 & 29 & 22 & 1.15 & 26 & 32 \tabularnewline
$20-40\thinspace (29.5)$ & 763 & 0.073 & 8.2 & 12 & 6.10 & 8.2 & 20 \tabularnewline
$40-60\thinspace (49.0)$ & 588 & 0.104 & 10 & 11 & 5.06 & 10 & 15 \tabularnewline
$60-200\thinspace (92.7)$ & 487 & 0.108 & 13 & 8.3 & 3.41 & 13 & 13 \tabularnewline
\hline 
\end{tabular}
\end{table}

\begin{figure*}
\centering
\includegraphics[width=0.45\textwidth]{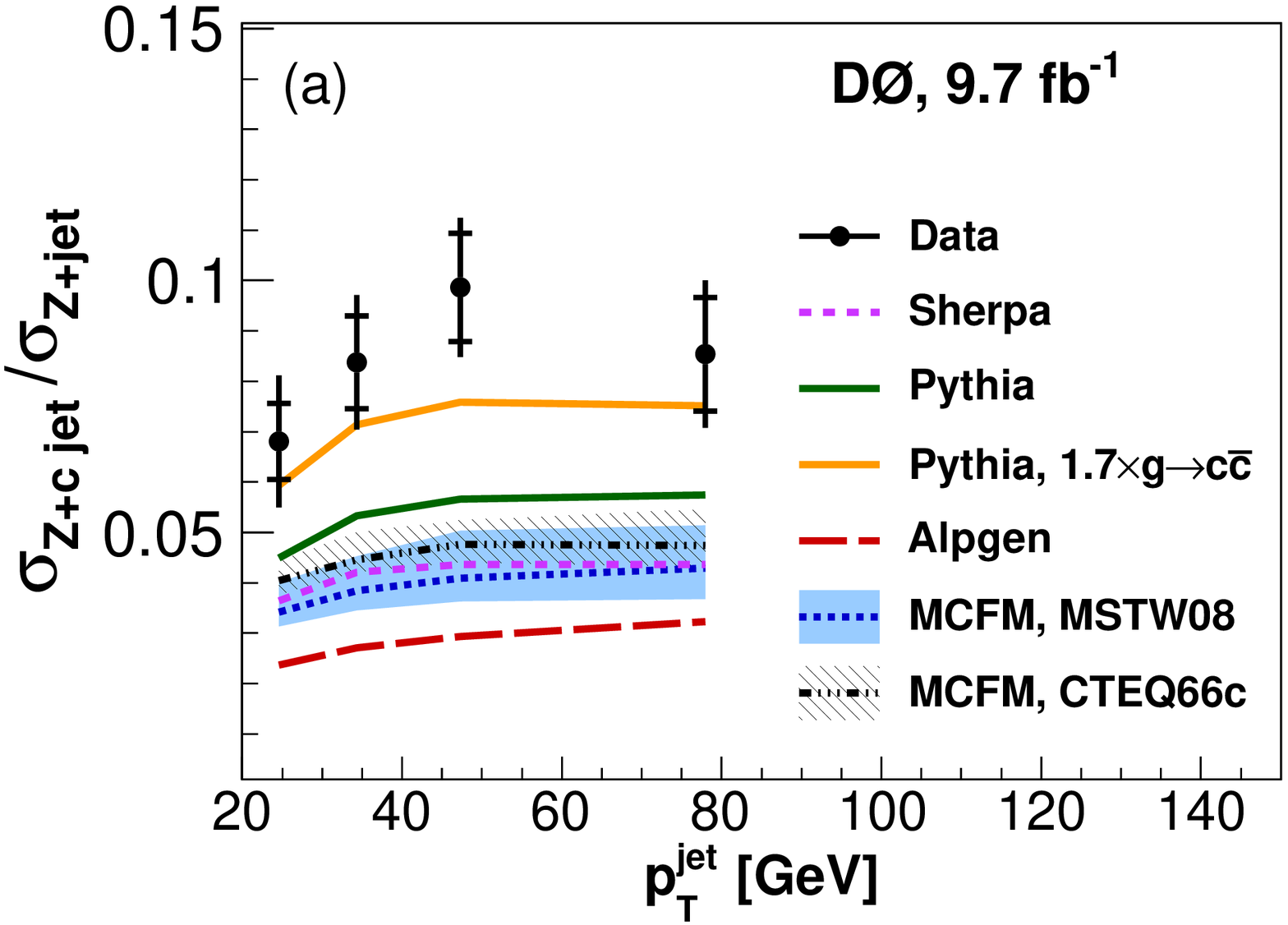}
\includegraphics[width =0.45\textwidth]{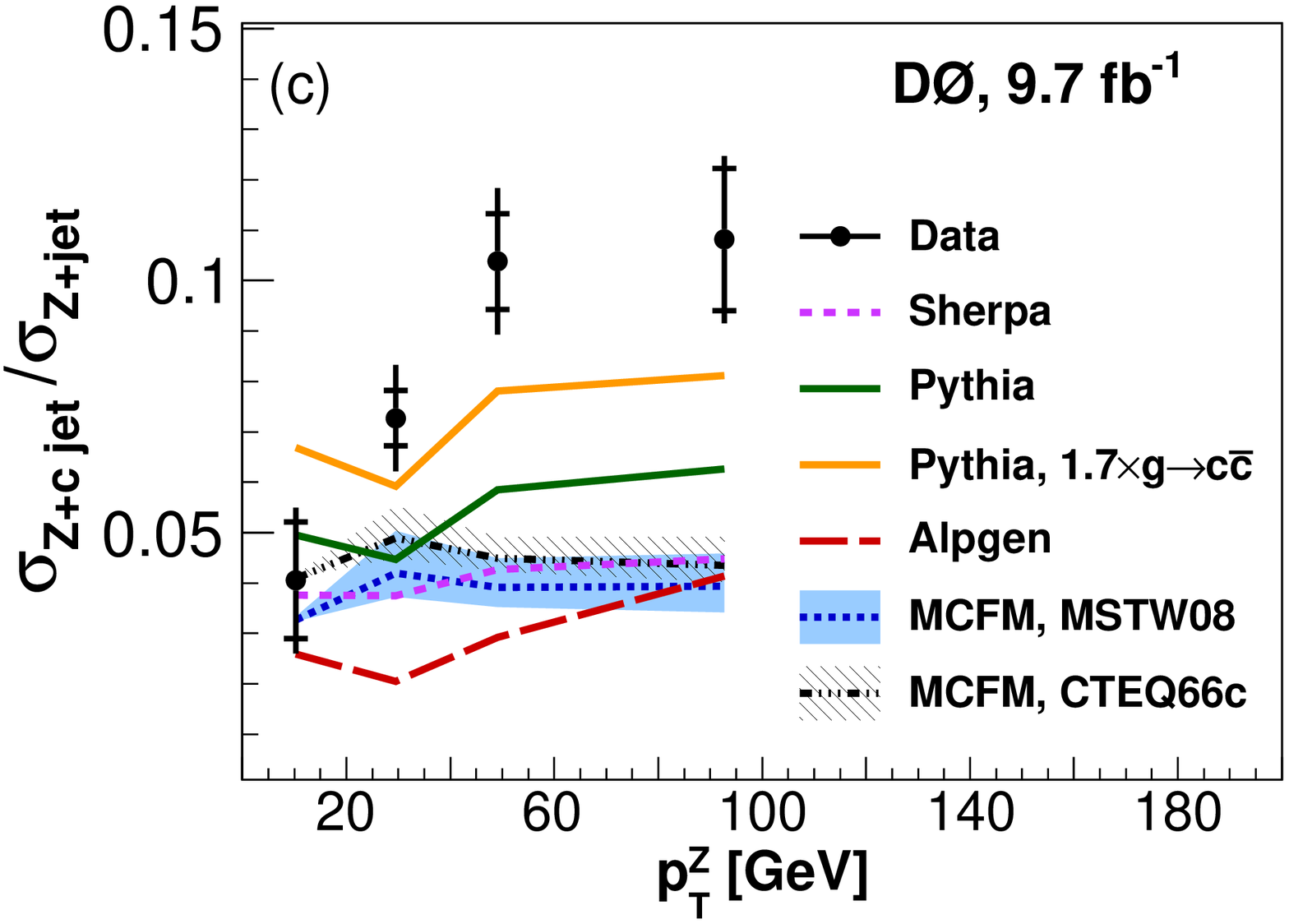}\\
\includegraphics[width =0.45\textwidth]{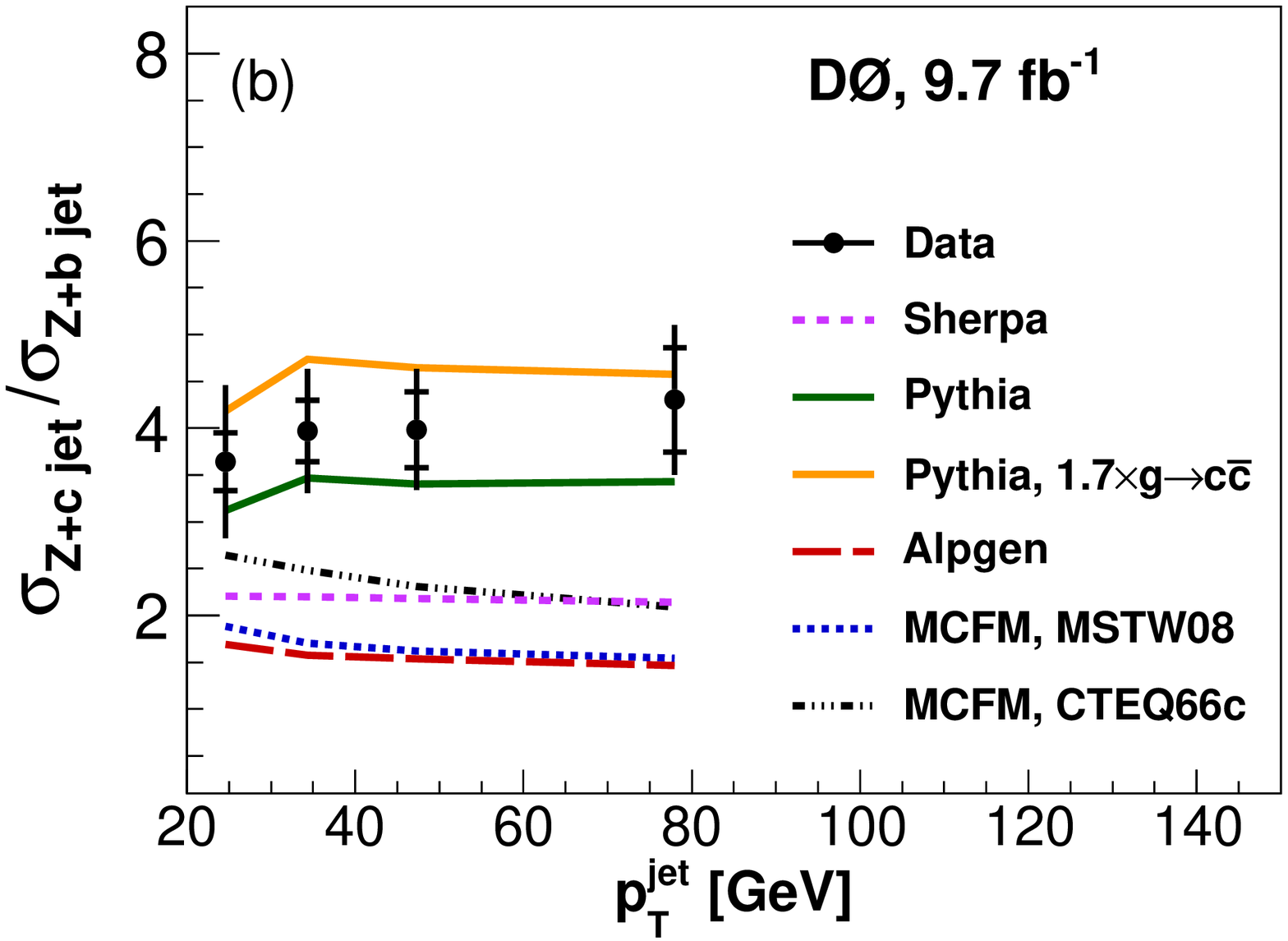}
\includegraphics[width =0.45\textwidth]{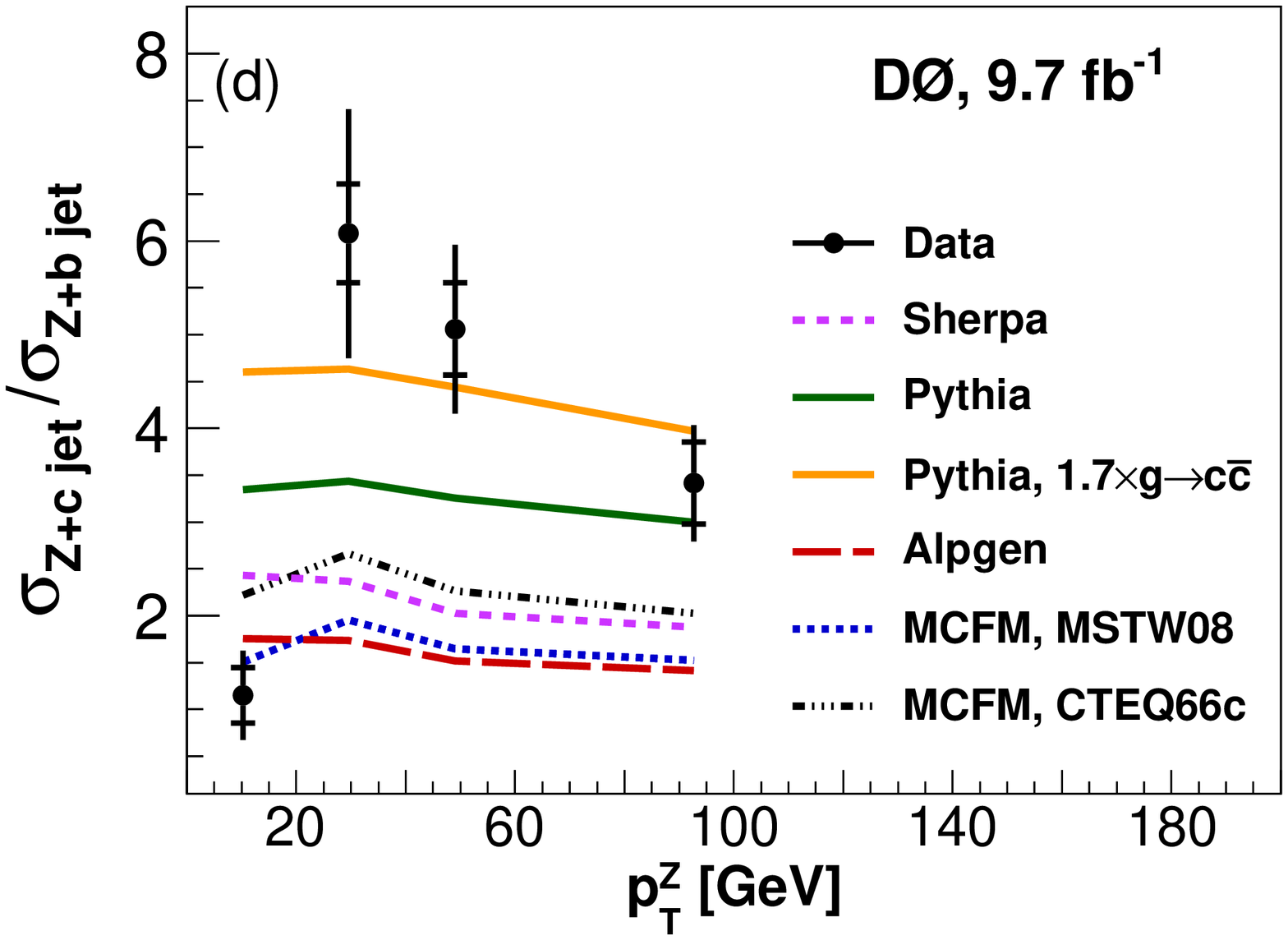}

\caption{\label{fig:diffs} (color online) Ratios of the differential cross sections 
\ZcZj~and \ZcZb~as a function of (a,b) \ptj~and (c,d)~\ptze, respectively. 
The uncertainties on the 
data include statistical (inner error bar) and full uncertainties (entire error bar).  
The predictions from {\sc alpgen}, \sherpae, \pythiae, \pythia with an enhanced $g \rightarrow c\bar{c}$~component,
and \mcfm NLO with the MSTW2008 and the {\sc cteq}6.6c PDFs are also shown.
The bands represent variations of the scales up and down by a factor of two.}
\end{figure*}

Several systematic uncertainties cancel when the ratios are measured. 
These include uncertainties
on the luminosity measurement, as well as trigger, lepton, and 
the jet reconstruction efficiencies. The remaining uncertainties 
are estimated separately for the integrated and differential results.
For the two ratios the systematic uncertainties are estimated separately.

For the integrated \ZcZj~measurement, the largest systematic uncertainty of 8.1\%
comes from the estimation of the $Z+\text{light jet}$ background. This 
is quantified by comparing the value extracted 
from the data, using a three template (light, $c$, and $b$ jet) fit, for various \bl selections.
The next largest systematic uncertainty comes from the 
shape of the \mjl templates used in the fit. 
A variety of different aspects can affect the shape of the templates: 
two HF jets being reconstructed as a single jet; 
models of $b$ and $c$ quark fragmentation;
the background from the non-$Z+\text{jet}$ events; 
the difference in the shape of the light jet MC template and a template derived from 
a light jet enriched dijet data sample; 
and the uncertainty of shape of the templates due to MC statistics. 
These are all evaluated by varying
the central values by the corresponding uncertainties, one at a time, and repeating
the entire analysis chain, resulting in a 5.5\% uncertainty.
An additional uncertainty of 3.4\% comes from jet energy calibration; 
it comprises the uncertainties on the jet energy 
resolution and the jet energy scale. 
An uncertainty is also associated with the $c$ jet tagging efficiency (1.9\%)~\cite{bid}. 
Finally, a small contribution ($<0.1\%$) is coming from the dependence of the acceptance on
modeling of the signal events. When summed in quadrature the
total systematic uncertainty for the integrated \ZcZj~ratio is 10.6\%.
The corresponding total systematic uncertainty is 14.4\% for \ZcZb. 
Table \ref{tab:Final} lists the total statistical and systematic uncertainties 
(added in quadrature) for the differential results.
Finally, for the integrated ratios we obtain values of
$\ZcZj = 0.0829\pm0.0052 \thinspace (\text{stat.})\pm0.0089 \thinspace (\text{syst.})$
and
$\ZcZb = 4.00\pm0.21 \thinspace (\text{stat.})\pm0.58 \thinspace (\text{syst.})$.  

The measurements are compared to predictions from an \mcfm NLO pQCD calculation
and three MC event generators, \sherpa \cite{Sherpa}, \pythiae, and \alpgene.
The NLO predictions are based on {\sc mcfm}~\cite{Campbell}, version 6.3,
with the MSTW2008 PDFs~\cite{mstw}
and the renormalization and factorization scales set at
$\mu_{R}^{2}=\mu_{F}^{2}=M_{Z}^{2}+p^{2}_{T\text{,total}}$. Here, $M_{Z}$ is 
the $Z$ boson mass and $p_{T\text{,total}}$~is the scalar sum of the transverse 
momentum for all the jets with \ptj$> 20$~GeV and $|\eta| < 2.5$ in the event. 
Corrections are applied to account for non-perturbative effects, 
on the order of $5\%$, estimated using the {\sc alpgen+pythia} simulation.
The NLO pQCD predictions of $\ZcZj = 0.0368$ and $\ZcZb = 1.64$~\cite{Campbell}
disagree significantly with the measurements. In the case where 
the intrinsic charm of the proton is enhanced, as suggested in the 
{\sc cteq}6.6c PDF sets~\cite{cteq6}, \mcfm yields ratios of 
$\ZcZj = 0.0425$
and
$\ZcZb = 2.23$, which are still in disagreement with our data.

The uncertainty on the \ZcZj~theoretical predictions are evaluated 
by simultaneously changing the $\mu_{R}$ and $\mu_{F}$ scales 
up and down by a factor of two, yielding an uncertainty of up to 11\% 
on \ZcZj, while this uncertainty cancels in \ZcZb. 
However, this uncertainty is smaller than the effect due to the 
intrinsic charm enhancement, which is 15\% and 
36\% for \ZcZj~and \ZcZb, respectively.

\alpgen generates multi-parton final states using tree-level
MEs. When interfaced with {\sc pythia}, it
employs the MLM scheme~\cite{MLM} to match ME 
partons with those after showering in {\sc pythia}, 
resulting in an improvement over leading-logarithmic
accuracy. 

\sherpa uses the CKKW matching scheme between the leading-order ME
partons and the parton-shower jets following the prescription given in
Ref.~\cite{CKKW}. This effectively allows for
a consistent combination of the ME and
parton shower.

\pythia includes only $2 \rightarrow 2$ MEs with $gQ \rightarrow ZQ$ and
$q\bar{q}\rightarrow Zg$ scatterings followed by $g \rightarrow Q\bar{Q}$
splitting, where $Q$ is either a $b$ or $c$ quark. The Perugia0 tune~\cite{perugia} 
and the {\sc cteq6l1} PDF set are used for the \pythia predictions. 

The ratios of differential cross sections as a function of 
\ptj and \ptz are compared to various predictions in Fig.~\ref{fig:diffs}.
On average, the NLO predictions significantly underestimate the data, 
by a factor of 2.5 for the integrated results.
As for the MC event generators, \pythia predictions are closer to data. 
An improved description can be achieved by enhancing 
the default rate of $g \rightarrow c\bar{c}$ in \pythia by a factor of 
1.7, motivated by the $\gamma + c~\text{jet}$ 
production measurements at the Tevatron~\cite{gammac,CDFgc}.

The largest discrepancy between data and predictions, in particular for the
shape of the differential distributions, is for \ZcZb~as
a function of \ptze~(Fig.~\ref{fig:diffs}(d)). 
The level of disagreement in shape is
quantified for the \mcfm NLO prediction when its integrated result is
scaled up to match the data. We generated a large number
of pseudo-experiments and found the p-value for the four bins in \ptz
to simultaneously fluctuate to the observed \ZcZb~values (or beyond) to be 2\%.

We have presented the first measurements of the ratios of integrated cross sections,
$\sigma(p\bar{p}\rightarrow Z+c~\text{jet})$/$\sigma(p\bar{p}\rightarrow Z+\text{jet})$ 
and $\sigma(p\bar{p}\rightarrow Z+c~\text{jet})$/$\sigma(p\bar{p}\rightarrow Z+b~\text{jet})$,
as well as the ratios of the differential cross sections in bins of
\ptj and \ptze, for events with a $Z$ boson decaying to electrons or muons 
and at least one jet in the final state. Measurements are based on
the data sample collected by the \dzero experiment in Run~II of the Tevatron,
corresponding to an integrated luminosity of 9.7~fb$^{-1}$ 
at a center-of-mass energy of 1.96 TeV. For jets with
\ptj~$>20$~GeV and $|$\etaj$|<2.5$, the measured integrated ratios are
$\ZcZj=0.0829\pm0.0052 \thinspace (\text{stat.})\pm0.0089 \thinspace (\text{syst.})$,
and  
$\ZcZb=4.00\pm0.21\thinspace (\text{stat.})\pm0.58 \thinspace (\text{syst.})$.
The NLO pQCD predictions disagree significantly with the results.  
\pythia agrees better with the measured ratios, especially when 
the gluon splitting to $c\bar{c}$ pairs is enhanced.

%
We thank the authors of Refs.~\cite{Campbell, Sherpa} for
valuable discussions, and the staffs at Fermilab, 
and collaborating institutions,
and acknowledge support from the
DOE and NSF (USA);
CEA and CNRS/IN2P3 (France);
MON, NRC KI and RFBR (Russia);
CNPq, FAPERJ, FAPESP and FUNDUNESP (Brazil);
DAE and DST (India);
Colciencias (Colombia);
CONACyT (Mexico);
NRF (Korea);
FOM (The Netherlands);
STFC and the Royal Society (United Kingdom);
MSMT and GACR (Czech Republic);
BMBF and DFG (Germany);
SFI (Ireland);
The Swedish Research Council (Sweden);
and
CAS and CNSF (China).


\begin{thebibliography}{10}
 \bibitem{Campbell} J.~M.~Campbell, R.~K.~Ellis, F.~Maltoni,
and S.~Willenbrock, Phys. Rev. D \textbf{69}, 074021 (2004). 

\bibitem{zhllbb} V.~M.~Abazov \textsl{et al.} (D0 Collaboration),
Phys. Rev. Lett. \textbf{109}, 121803 (2012);
T.~Aaltonen \textsl{et al.} (CDF Collaboration), 
Phys. Rev. Lett. \textbf{109}, 111803 (2012).

\bibitem{CDFPaper} A.~Abulencia \textsl{et al.}
(CDF Collaboration), Phys. Rev. D \textbf{74}, 032008 (2006).

\bibitem{CDFPaperII} T.~Aaltonen \textsl{et al.}
(CDF Collaboration), Phys. Rev. D \textbf{79}, 052008 (2009). 

\bibitem{Zb_PRL} V.~M.~Abazov \textsl{et al.} (D0 Collaboration), 
Phys. Rev. Lett. \textbf{94}, 161801 (2005).

\bibitem{Zb} V.~M.~Abazov {\it et al.}  (D0 Collaboration),  Phys. Rev. D {\bf 87}, 092010 (2013).

 \bibitem{d0det}  V.M.~Abazov {\sl et al.} (D0 Collaboration),
Nucl. Instrum. Methods Phys. Res. A {\bf 565}, 463  (2006);
M.~Abolins {\sl et al.}, Nucl. Instrum. Methods Phys. Res. A {\bf 584}, 75  (2008);
R.~Angstadt {\sl et al.}, Nucl. Instrum. Methods Phys. Res. A {\bf 622}, 298  (2010).

\bibitem{def} Pseudorapidity is defined as $\eta=-\ln[\tan(\theta/2)]$, with the polar
angle $\theta$ measured relative to the proton beam direction.

\bibitem{RunIIcone} G.~C.~Blazey \textsl{et al.}, arXiv:hep-ex/0005012.

\bibitem{alpgen} M.~L.~Mangano \textsl{et al.}, J. High Energy Phys.
\textbf{07} (2003) 001. Version 2.11 was used. 

\bibitem{pythia} T.~Sj\"{o}strand, S.~Mrenna, and P.~Skands, J. High
Energy Phys. \textbf{05} (2006) 026. Version 6.409 was used.

\bibitem{evtgen} D. J. Lange, 
Nucl. Instrum. Meth. Phys. Res. A \textbf{462}, 152 (2001).

\bibitem{cteq6} J.~Pumplin \textsl{et al.}, J. High Energy Phys.
\textbf{07} (2002) 012. 

\bibitem{diboson} J.~M.~Campbell and R.~K.~Ellis, Phys. Rev. D \textbf{60},
113006 (1999); ibid. \textbf{62}, 114012 (2000); ibid. \textbf{65}, 113007 (2002).

\bibitem{ttbar} U.~Langenfeld, S.~Moch, and P.~Uwer,  Phys. Rev. D \textbf{80},
054009 (2009).

\bibitem{zpt} V.~M.~Abazov \textsl{et al.} (D0 Collaboration),
Phys. Rev. Lett. \textbf{100}, 102002 (2008).

\bibitem{bid} V.~M.~Abazov \textsl{et al.} (D0 Collaboration),
Nucl. Instrum. Methods Phys. Res. Sect. A \textbf{620}, 490 (2010).
V.~M.~Abazov \textsl{et al.} (D0 Collaboration), to be submitted to Nucl. Instrum. Methods Phys. Res. A.

\bibitem{MVA} A.~Hoecker, P.~Speckmayer, J.~Stelzer, J.~Therhaag, E.~von Toerne, and H.~Voss,
``TMVA: Toolkit for Multivariate Data Analysis,'' PoS A CAT 040 (2007) [physics/0703039].

\bibitem{Terry} G.D.~Lafferty, and T.R.~Wyatt, Nucl. Instrum. Methods Phys. Res. Sect. A \textbf{355}, 541 (1995).

\bibitem{Sherpa} T.~Gleisberg \textsl{et al.}, J. High Energy Phys. \textbf{02} (2009) 007.

\bibitem{mstw} A.~D.~Martin, W.~J.~Stirling, R.~S.~Thorne, 
and G.~Watt, Eur. Phys. J. C \textbf{63}, 189 (2009).

\bibitem{MLM} F.~Caravaglios \textsl{et al.}, Nucl. Phys. \textbf{B539}, 215 (1999).

\bibitem{CKKW} S.~Catani \textsl{et al.}, J. High Energy Phys. \textbf{11} (2001) 063. Version 1.3.1 was used.

\bibitem{perugia} P.~Z.~Skands,  Phys.\ Rev.\ D {\bf 82}, 074018 (2010).

\bibitem{gammac}  V.~M.~Abazov {\it et al.}  (D0 Collaboration),  Phys.\ Lett.\ B {\bf 719}, 354 (2013).

\bibitem{CDFgc} T.~Aaltonen {\it et al.}  (CDF Collaboration), 
Phys. Rev. Lett. {\bf 111}, 042003 (2013).

\end{thebibliography}
\end{document}